\documentclass[submission,copyright,creativecommons]{eptcs}
\providecommand{\event}{Machines, Computations and Universality 2013} % Name of the event you are submitting to

\usepackage{breakurl}
\usepackage{amsthm}
\usepackage[utf8]{inputenc}
\usepackage[english]{babel}
\usepackage[]{hyperref} % colorlinks
\usepackage{microtype}   % typesetting improvements to tex
\usepackage{graphicx}

\DeclareMathAlphabet{\mathcal}{OMS}{cmsy}{m}{n}

\begin{document}

\title{Intrinsic universality and  the computational power of self-assembly}
\author{Damien Woods\thanks{Supported by National Science Foundation grants 0832824 (The Molecular Programming Project), CCF-1219274, and CCF-1162589, and NASA grant number NNX13AJ56G.}   
\institute{California Institute of Technology  
\email{woods@caltech.edu}}}
\def\titlerunning{Intrinsic universality and the computational power of self-assembly}
\def\authorrunning{Damien Woods}

\date{}

\maketitle

\begin{abstract}
This short survey of recent work in tile self-assembly discusses the use of simulation to classify and separate the computational and expressive power of self-assembly models.  The journey begins with the result that there is a single universal tile set that, with proper initialization and scaling, simulates any tile assembly system.  This universal tile set exhibits something stronger than Turing universality: it captures the geometry and dynamics of any simulated system.  From there we find that there is no such tile set in the noncooperative, or temperature~1, model, proving it weaker than the full tile assembly model.  In the two-handed or hierarchal model, where large assemblies can bind together on one step, we encounter an infinite set, of infinite hierarchies, each with strictly increasing simulation power.  Towards the end of our trip, we find one tile to rule them all: a single rotatable flipable polygonal tile that can simulate any tile assembly system.  It seems this could be the beginning of a much longer journey, so directions for future work are suggested.
\end{abstract}

\subsection*{Introduction}
Self-assembly is the process by which  small pieces of matter coalesce according to simple rules to form some kind of target structure. Theoretical models of self-assembly have been used to design a variety of  DNA nanoscale structures that have been experimentally implemented, and indeed many view it as imperative to have a theory of self-assembly to guide this rapidly developing engineering field.  Self-assembly models are capable of Turing-universality, and so of an infinite variety of interesting computational behaviours.

The abstract Tile Assembly Model, put forward by Winfree~\cite{Winf98}, is one such model. It is a kind of  asynchronous nondeterministic cellular automaton that models crystal growth processes.  Put another way, the abstract Tile Assembly Model  restricts classical square Wang tiling \cite{Wang61} to use a mechanism for crystal-like growth of a tiling, one tile at a time, starting from a special seed tile. Of course individual tiles need not be square, indeed triangles~\cite{kari2012triangular}, hexagons~\cite{nubots,one,kari2012triangular} and arbitrary  polygons have been considered~\cite{one}. Other models of self-assembly allow for non-local rules and large-scale interactions. One such model is the two-handed (or hierarchical) model~\cite{Versus,2HAMIU} which allows large structures to stick together if enough of their tiles' edge colours  match. Another  is the Nubots model~\cite{nubots} where large assemblies of molecules can grow and then move relative to each other, implementing a kind of molecular robotics. Although some forms of self-assembly can be thought of as generalizations of cellular automata, or  effectivisations of Wang tiling, these models are quite distinct from each other both in terms of questions that can be asked and results that can be obtained.

\begin{figure}[p]
    \centering
% uncomment the following line to run tikz source code
% \input{simulation_classes}
    \includegraphics[width=0.97\textwidth]{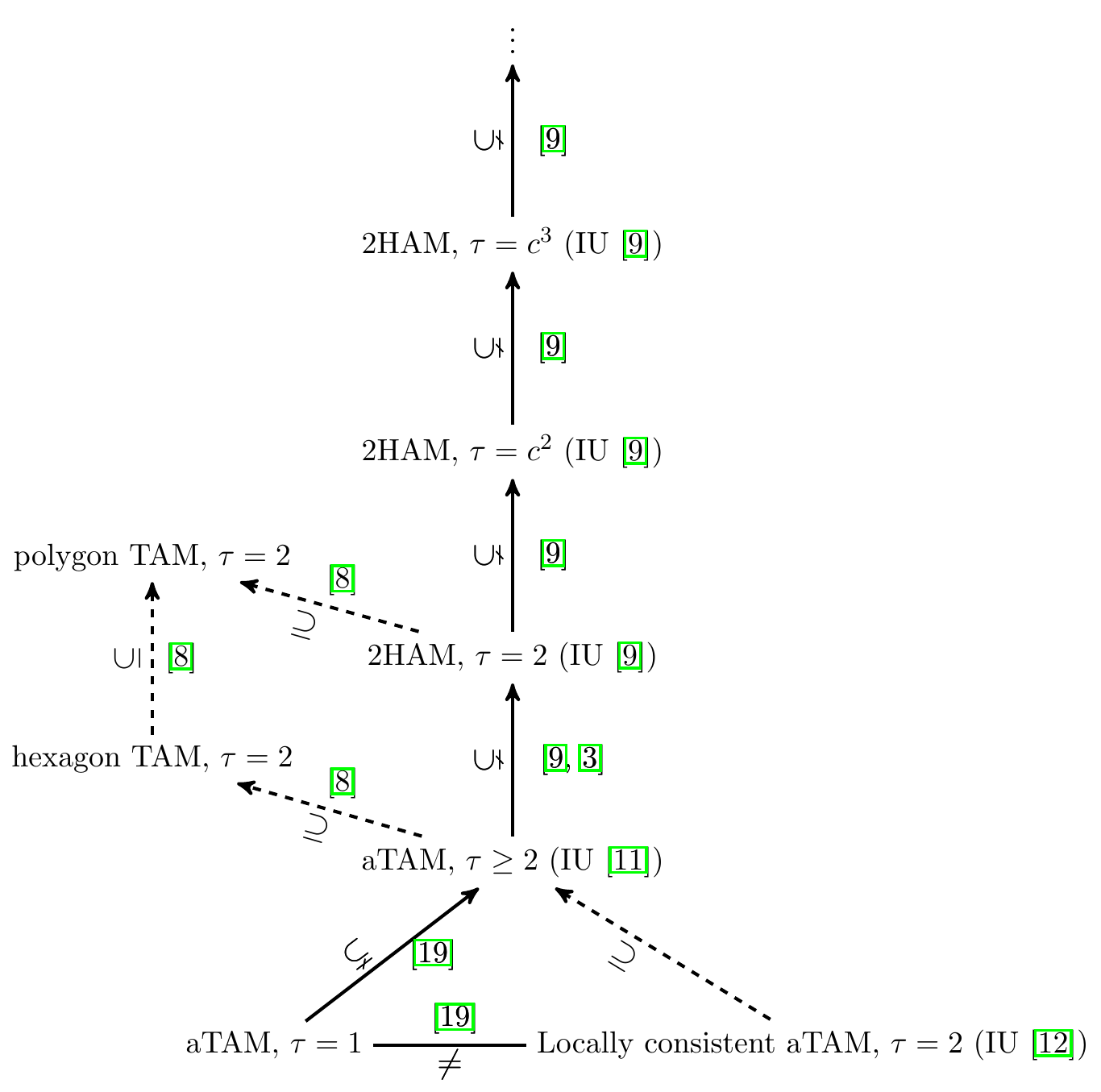}
    %\vspace{-4ex}
    \caption{Classes of tile assembly systems, and their relationship with respect to simulation. aTAM: abstract Tile Assembly Model (growth from a seed assembly by single tile addition in 2D), $\tau$ denotes ``temperature''.  2HAM: Two-Handed Tile Assembly Model (individual tiles and large assemblies  stick together in 2D). Class $A$ contains  class $B$ if there is a tile set $U$ that simulates any member of $B$ and  does so in a tile assembly system from $A$. Solid arrows denote strict containment, dashed arrows denote containment. A 2HAM temperature hierarchy is shown for some $c \in \{2,3,4,\ldots \} $. In fact, for each such $c$  the set $\{c^i | i \in \{ 1,2,3,\ldots \} \}$ gives an infinite hierarchy of classes of strictly increasing simulation power in the 2HAM.  The letters ``IU'' denote the existence of an intrinsically universal tile set for a class, and omission of ``IU'' implies that the existence of such a tile set is an open problem for that class. Citations are given in square brackets. Simulation results for a number of other models are described in the main text.}
    \label{fig:simulationclasses}
  \end{figure}

 Here we discuss the relative computational and expressive power of self-assembly models, with much of the discussion summarized in Figure~\ref{fig:simulationclasses}.  Specifically, we seek to classify tile assembly models based upon their ability to simulate each other, or not.  To do this, tile assemblers have borrowed a powerful idea from the cellular automata community: intrinsic universality.  The topic of intrinsic universality, with its tight notion of simulation, has given rise to a rich theory in cellular automata~\cite{bulkingI,bulkingII,Ollinger08,ollinger-fourstates,goles-communicationcomplexity,arrighi2012intrinsic} and Wang tiling~\cite{LafitteW07,LafitteW09}. This short survey attempts to show that we are beginning to see this in  self-assembly too~\cite{IUSA,USA, one, 2HAMIU,  temp1notIU,hendricks2013signal,HendricksPatitzTAMCA}. 
  
\subsection*{The abstract Tile Assembly Model is intrinsically universal}
An instance of the abstract Tile Assembly Model~\cite{Winf98} is called a tile assembly system and is a triple consisting of a set of square tiles, a seed assembly (one or more tiles stuck together), and a temperature $\tau \in \{ 1,2,3,\ldots \} $. Each side of a square tile has a glue (or colour) $g$ and strength $s \in \{ 0,1,2, \ldots \}$. Growth begins from the seed assembly.  A tile sticks to a partially-formed assembly if it can be positioned  so that   enough of  its glues match those of adjacent tiles on the assembly and the sum of the matching glue strengths  exceeds the temperature. Growth  proceeds one tile at a time, asynchronously and nondeterministicly.  (Tiles may not overlap nor rotate, and  unlike Wang tiling, adjacent glues (colours) may mismatch in an assembly.) This is a formal model of a  crystal-like growth process.
  
Recently, the abstract Tile Assembly Model has been shown to be intrinsically universal~\cite{IUSA}, meaning that there is a single set of tiles~$U$ that when appropriately initialized, is capable of simulating the behavior of an arbitrary  tile assembly system $\mathcal{T}$, up to rescaling. Specifically, this means that (1)~each tile of $\mathcal{T}$ is represented by a ``supertile'' of $U$, that is, by a $k\times k$  block of tiles (where $k$  depends on the size of $\mathcal{T}$'s tile set, but is independent of what~$\mathcal{T}$ does),
(2)~the simulated system~$\mathcal{T}$ is encoded  in a $k \times k$ ``seed supertile'' of the simulator, and (3)~every sequence of tile placements in the simulated system is simulated by a sequence of supertile placements in the universal simulator system, and vice-versa. In other words, this intrinsically universal tile set has the ability to simulate both the geometry and growth order of any tile assembly system. Modulo rescaling, this universal tile set $U$ represents the full power and expressivity of the entire abstract Tile Assembly Model. 

\subsection*{Temperature~1: negative results in intrinsic universality}
It has been known for some time that the temperature~2 tile assembly model, where at least  {\em  some} of the tiles are  required to match on {\em two or more} sides,  i.e.\  {\em cooperative growth}, leads to highly non-trivial behavior.   Turing machine simulation~\cite{RotWin00}, efficient production of $n \times n$ squares and other simple shapes using $\Theta(\log n/\log \log n)$ tile types~\cite{AdChGoHu01}, efficient production of arbitrary finite connected shapes using a number of tile types  within a log factor of the Kolmogorov complexity of the shape~\cite{SolWin07}, and even intrinsic universality~\cite{IUSA} (as mentioned above) can all be achieved with cooperative, or temperature~2, growth.

The fact that the abstract Tile Assembly Model is intrinsically universal suggests an obvious question: can we show that some subclasses of the model are provably weaker than the full model, by showing that these subclasses are not universal for the full model? 

The most notorious such subclass is called temperature~1. Despite its esoteric name, it models a fundamental and ubiquitous form of growth:  asynchronous growing and branching tips in Euclidian space where each new tile is added if matches on at {\em least one side}. Since temperature~1 binding requires matching glues on only one or more sides, it is called noncooperative binding. In 2D, its like snakes on a plane.  

Recently, ideas from intrinsic universality have been used to show that the Temperature~1  Tile Assembly Model (i.e.\ noncooperative binding) is provably weaker than the full tile assembly model~\cite{temp1notIU}: in particular, temperature~1  tile assembly is not capable of {\em simulating} arbitrary tile assembly systems.  In fact, there is a very simple cooperative tile assembly system that can not be simulated by any noncooperative tile assembly system.  This is the first fully general negative result about temperature~1 that does not assume restrictions on the model nor unproven hypotheses. (However, it  tantalizingly leaves open the question of whether 2D deterministic temperature~1 systems are Turing universal, or whether  they can efficiently assemble  $n \times n$ squares with less than $2n-1$ tile types.) An interesting aspect of this negative result is that it holds for 3D noncooperative systems, they too can not simulate arbitrary tile assembly systems. This is quite shocking, given that 3D noncooperative systems are Turing-universal~\cite{Cook-2011}! Hence, the negative result in~\cite{temp1notIU} can be interpreted to mean that Turing-universal algorithmic behavior in self-assembly does not imply the ability to simulate arbitrary  algorithmic self-assembly processes. 

What about other subclasses of the abstract tile assembly  model? It was shown that 3D noncooperative systems can simulate 2D noncooperative systems~\cite{temp1notIU}. The same paper contains the conjecture that there is no 2D noncooperative tile set that can simulate all 2D noncooperative tile assembly systems. 

The first intrinsic universality result in tile assembly showed that locally consistent tile assembly systems (a restriction of the general model where binding uses at most two sides, with bonds that are not ``too strong'') have an intrinsically universal tile set~\cite{USA}. It turns out that even these locally consistent systems can not be simulated by noncooperative systems~\cite{temp1notIU}. As observed in~\cite{USA}, Soloveichik and Winfree's construction~\cite{SolWin07} gives a tile set that, when suitably seeded, simulates any temperature~1 tile assembly system whose underlying binding graph is a spanning tree.

It remains as future work to further characterize the power of interesting subclasses of the abstract Tile Assembly Model, and in particular, to separate such subclasses.

\subsection*{Simulation and intrinsic universality in other models of self-assembly }
What about other models of self assembly besides the abstract Tile Assembly Model? It has been shown that  the two-handed, or hierarchical, model of self-assembly (where large assemblies of tiles may come together in a single step) is not intrinsically universal~\cite{2HAMIU}. Specifically there is no tile set that can simulate all two-handed systems for all temperatures. However, the same paper shows that for each temperature $\tau \in \{ 2,3,4, \ldots \}$ there is a tileset that is intrinsically universal for the class of two-handed systems that work at temperature~$\tau$. Also, there is an infinite hierarchy of classes of such systems with each level strictly more powerful than the one below. In fact there are an infinite set of such hierarchies, as described in the caption of Figure~\ref{fig:simulationclasses}. These results~\cite{2HAMIU} formalize the intuition that multiple long range interactions are more powerful than  fewer long range interactions in the two-handed model. 

By combining results from~\cite{Versus} and~\cite{IUSA}, we get that there is a tile set for the two-handed model that (at temperature~2) simulates any instance (assembly system) of the abstract Tile Assembly Model, as shown in  Figure~\ref{fig:simulationclasses}.

The Signal-passing Tile Assembly Model has tiles with molecular (DNA) wires on their surface. Essentially there is a  ``use once'' circuit sitting on the assembly itself!  These wires use an experimentally plausible DNA strand displacement mechanism.  Recently, Hendricks, Padilla, Patitz and Rogers~\cite{hendricks2013signal} have shown  that in the two-handed model there is a single 3D  tile set, that can simulate any 2D signal-passing tile assembly system. Their result shows that a constant number of planes (in the third dimension) is sufficient to handle wire crossings and asynchronous signal passing. 

As mentioned above, the Tile Assembly Model can be thought of as an asynchronous and nondeterministic cellular automaton (CA), that models the notion of a crystal growth frontier. Hendricks and Patitz~\cite{HendricksPatitzTAMCA} formally relate the abstract Tile Assembly Model and CA: they give a single CA that simulates any tile assembly system, as well as a single tile set that simulates any  nondeterministic CA with a finite initial configuration. The methods of updating configurations in both models are quite different (CA are infinite, synchronous and deterministic, while tile assembly is finite, asynchronous and nondeterministic), and so their constructions need to handle this. Their  work opens the possibility of further comparing and contrasting the CA and tile assembly models.

\subsection*{The One}
Demaine, Demaine, Fekete, Patitz, Schweller,  Winslow, and Woods~\cite{one} take the existence of an  intrinsically universal tile set for the abstract Tile Assembly Model~\cite{IUSA} as merely the first  in a sequence of simulations that routes from squares tiles, to the intrinsically universal tile set, to (low-strength) hexagons to a {\em single}  polygon that is translatable, rotatable and flipable. Their   fixed-sized polygon, when appropriately seeded,  simulates any tile assembly system. This polygon, {\em The One}, captures the power of the entire abstract Tile Assembly Model: to simulate a tile assembly system $\mathcal{T}$ one simply puts together a seed assembly of polygons that encodes $\mathcal{T}$ and just lets it go! Likewise, Turing machines can be simulated with this single tile. They also show~\cite{one} that with translation only, such results are not possible with a small (size $\leq 3$) seed. In the simpler setting of Wang plane tilling, they show~\cite{one} how to take any  tile set $T$ (on the square or hexagonal lattice) and ``compile'' it to get a single regular polygon that simulates exactly the tilings of~$T$, except with tiny gaps between the polygons. In particular, if one starts with any aperiodic square or hexagon tile set, that set can be easily complied to a {\em single regular polygonal tile}, all of who's tilings are aperiodic, with tiny gaps between the polygons.

\subsection*{Future work}
As Figure~\ref{fig:simulationclasses} shows, we are beginning to classify the power of tile assembly models using intrinsic universality. Indeed gaps in the figure suggest obvious future directions.  It is  an open question whether or not the hexagonal Tile Assembly Model~\cite{one}, polygonal Tile Assembly Model~\cite{one}, Nubots model~\cite{nubots} and  Signal-passing Tile Assembly Model~\cite{hendricks2013signal,jonoska2012active} are intrinsically universal. And independent of whether or not these models turn out to be intrinsic universal, we suggest that simulation can be used to tease apart their computational and expressive power, as well as the power of subclasses of these models. 

A more technical question: Does there exist a tile set $U$ for the aTAM, such that for any (adversarially chosen) seed assembly~$\sigma$, at temperature~2, this tile assembly system  simulates some tile assembly system~$\mathcal{T}$? Moreover,~$U$ should be able to simulate all such members $\mathcal{T}$ of some (hopefully, non-trivial) class~$S$. $U$~is a tile set that can do one thing and nothing else: simulate tile assembly systems from $S$. This question about $U$ is inspired by the factor simulation question in CA~\cite{bulkingII}, although it differs  in the details. 

Many cooperative ``algorithmic'' tile assembly systems (temperature~2 and above) simulate Turing machines in a ``zig-zag'' fashion. Can the technique of~\cite{temp1notIU} be extended to show that no 2D temperature~1 aTAM system simulates   zig-zag aTAM tile assembly systems? This would be a non-trivial extension as~\cite{temp1notIU} holds for the 3D temperature~1 aTAM, which can indeed simulate zig-zag systems, and would show that no deterministic 2D temperature~1 system can simulate Turing machines in the ``usual way''.

Of course, there are many other ways to compare the power of self-assembly models~\cite{patitz2013introduction,dotysurvey}: researchers have looked at shape and pattern building, tile complexity, time complexity,  determinism versus nondeterminism, and randomization in self-assembly, and it remains as important future work to find relationships between these notions on the one hand, and intrinsic universality and simulation on the other hand. Can ideas from intrinsic universality be used to answer questions about these notions? 

\section*{Acknowledgements} 
I would like to thank my co-authors on the topic of intrinsic universality in tile assembly. It's been a fun ride so far, yet it seems there is much to learn.  I also thank Erik Winfree and Paul W.K. Rothemund for  discussions on intrinsic universality over the years, and Pierre-\'Etienne Meunier  and Matt Patitz for helpful comments on this paper.  Many thanks to Turlough Neary and Matt Cook for the invitation to talk at MCU~2013.

\bibliographystyle{eptcs}
\bibliography{t1IU} 

\begin{thebibliography}{10}
\providecommand{\bibitemdeclare}[2]{}
\providecommand{\surnamestart}{}
\providecommand{\surnameend}{}
\providecommand{\urlprefix}{Available at }
\providecommand{\url}[1]{\texttt{#1}}
\providecommand{\href}[2]{\texttt{#2}}
\providecommand{\urlalt}[2]{\href{#1}{#2}}
\providecommand{\doi}[1]{doi:\urlalt{http://dx.doi.org/#1}{#1}}
\providecommand{\bibinfo}[2]{#2}

\bibitemdeclare{inproceedings}{AdChGoHu01}
\bibitem{AdChGoHu01}
\bibinfo{author}{Leonard \surnamestart Adleman\surnameend},
  \bibinfo{author}{Qi~\surnamestart Cheng\surnameend}, \bibinfo{author}{Ashish
  \surnamestart Goel\surnameend} \& \bibinfo{author}{Ming-Deh \surnamestart
  Huang\surnameend} (\bibinfo{year}{2001}): \emph{\bibinfo{title}{Running time
  and program size for self-assembled squares}}.
\newblock In: {\sl \bibinfo{booktitle}{Proceedings of the 33rd Annual ACM
  Symposium on Theory of Computing}}, \bibinfo{address}{Hersonissos, Greece},
  pp. \bibinfo{pages}{740--748}, \doi{10.1145/380752.380881}.

\bibitemdeclare{inproceedings}{arrighi2012intrinsic}
\bibitem{arrighi2012intrinsic}
\bibinfo{author}{Pablo \surnamestart Arrighi\surnameend},
  \bibinfo{author}{Nicolas \surnamestart Schabanel\surnameend} \&
  \bibinfo{author}{Guillaume \surnamestart Theyssier\surnameend}
  (\bibinfo{year}{2012}): \emph{\bibinfo{title}{Intrinsic Simulations between
  Stochastic Cellular Automata}}.
\newblock In: {\sl \bibinfo{booktitle}{JAC 2012: 3rd international symposium
  Journ\'{e}es Automates Cellulaires}}, {\sl \bibinfo{series}{Electronic
  Proceedings in Theoretical Computer Science}}~\bibinfo{volume}{90},
  \bibinfo{publisher}{Open Publishing Association}, pp.
  \bibinfo{pages}{208--224}, \doi{10.4204/EPTCS.90.17}.

\bibitemdeclare{inproceedings}{Versus}
\bibitem{Versus}
\bibinfo{author}{Sarah \surnamestart Cannon\surnameend},
  \bibinfo{author}{Erik~D. \surnamestart Demaine\surnameend},
  \bibinfo{author}{Martin~L. \surnamestart Demaine\surnameend},
  \bibinfo{author}{Sarah \surnamestart Eisenstat\surnameend},
  \bibinfo{author}{Matthew~J. \surnamestart Patitz\surnameend},
  \bibinfo{author}{Robert \surnamestart Schweller\surnameend},
  \bibinfo{author}{Scott~M. \surnamestart Summers\surnameend} \&
  \bibinfo{author}{Andrew \surnamestart Winslow\surnameend}
  (\bibinfo{year}{2013}): \emph{\bibinfo{title}{Two Hands Are Better Than One
  (up to constant factors): Self-Assembly In The {2HAM} vs.\ {aTAM}}}.
\newblock In: {\sl \bibinfo{booktitle}{STACS: Proceedings of the Thirtieth
  International Symposium on Theoretical Aspects of Computer Science}}, {\sl
  \bibinfo{series}{LIPIcs}}~\bibinfo{volume}{20}, pp.
  \bibinfo{pages}{172--184}, \doi{10.4230/LIPIcs.STACS.2013.172}.

\bibitemdeclare{article}{goles-communicationcomplexity}
\bibitem{goles-communicationcomplexity}
\bibinfo{author}{Eric~Goles \surnamestart Ch.\surnameend},
  \bibinfo{author}{Pierre-Etienne \surnamestart Meunier\surnameend},
  \bibinfo{author}{Ivan \surnamestart Rapaport\surnameend} \&
  \bibinfo{author}{Guillaume \surnamestart Theyssier\surnameend}
  (\bibinfo{year}{2011}): \emph{\bibinfo{title}{Communication complexity and
  intrinsic universality in cellular automata}}.
\newblock {\sl \bibinfo{journal}{Theoretical Computer Science}}
  \bibinfo{volume}{412}(\bibinfo{number}{1-2}), pp. \bibinfo{pages}{2--21},
  \doi{10.1016/j.tcs.2010.10.005}.

\bibitemdeclare{inproceedings}{Cook-2011}
\bibitem{Cook-2011}
\bibinfo{author}{Matthew \surnamestart Cook\surnameend},
  \bibinfo{author}{Yunhui \surnamestart Fu\surnameend} \&
  \bibinfo{author}{Robert~T. \surnamestart Schweller\surnameend}
  (\bibinfo{year}{2011}): \emph{\bibinfo{title}{Temperature~1 self-assembly:
  deterministic assembly in~{3D} and probabilistic assembly in~{2D}}}.
\newblock In: {\sl \bibinfo{booktitle}{Proceedings of the 22nd Annual ACM-SIAM
  Symposium on Discrete Algorithms}}, pp. \bibinfo{pages}{570--589}.

\bibitemdeclare{article}{bulkingI}
\bibitem{bulkingI}
\bibinfo{author}{Marianne \surnamestart Delorme\surnameend},
  \bibinfo{author}{Jacques \surnamestart Mazoyer\surnameend},
  \bibinfo{author}{Nicolas \surnamestart Ollinger\surnameend} \&
  \bibinfo{author}{Guillaume \surnamestart Theyssier\surnameend}
  (\bibinfo{year}{2011}): \emph{\bibinfo{title}{Bulking {I}: an abstract theory
  of bulking}}.
\newblock {\sl \bibinfo{journal}{Theoretical Computer Science}}
  \bibinfo{volume}{412}(\bibinfo{number}{30}), pp. \bibinfo{pages}{3866--3880},
  \doi{10.1016/j.tcs.2011.02.023}.

\bibitemdeclare{article}{bulkingII}
\bibitem{bulkingII}
\bibinfo{author}{Marianne \surnamestart Delorme\surnameend},
  \bibinfo{author}{Jacques \surnamestart Mazoyer\surnameend},
  \bibinfo{author}{Nicolas \surnamestart Ollinger\surnameend} \&
  \bibinfo{author}{Guillaume \surnamestart Theyssier\surnameend}
  (\bibinfo{year}{2011}): \emph{\bibinfo{title}{{B}ulking {II}:
  {C}lassifications of cellular automata}}.
\newblock {\sl \bibinfo{journal}{Theoretical Computer Science}}
  \bibinfo{volume}{412}(\bibinfo{number}{30}), pp. \bibinfo{pages}{3881--3905},
  \doi{10.1016/j.tcs.2011.02.024}.

\bibitemdeclare{techreport}{one}
\bibitem{one}
\bibinfo{author}{Erik~D. \surnamestart Demaine\surnameend},
  \bibinfo{author}{Martin~L. \surnamestart Demaine\surnameend},
  \bibinfo{author}{S\'andor~P. \surnamestart Fekete\surnameend},
  \bibinfo{author}{Matthew~J. \surnamestart Patitz\surnameend},
  \bibinfo{author}{Robert~T. \surnamestart Schweller\surnameend},
  \bibinfo{author}{Andrew \surnamestart Winslow\surnameend} \&
  \bibinfo{author}{Damien \surnamestart Woods\surnameend}
  (\bibinfo{year}{2012}): \emph{\bibinfo{title}{One tile to rule them all:
  simulating any {T}uring machine, tile assembly system, or tiling system with
  a single puzzle piece}}.
\newblock \bibinfo{type}{Technical Report}.
\newblock \bibinfo{note}{Arxiv preprint
  \href{http://arxiv.org/abs/1212.4756}{\texttt{arXiv:1212.4756}} [cs.DS]}.

\bibitemdeclare{inproceedings}{2HAMIU}
\bibitem{2HAMIU}
\bibinfo{author}{Erik~D. \surnamestart Demaine\surnameend},
  \bibinfo{author}{Matthew~J. \surnamestart Patitz\surnameend},
  \bibinfo{author}{Trent~A. \surnamestart Rogers\surnameend},
  \bibinfo{author}{Robert~T. \surnamestart Schweller\surnameend},
  \bibinfo{author}{Scott~M. \surnamestart Summers\surnameend} \&
  \bibinfo{author}{Damien \surnamestart Woods\surnameend}
  (\bibinfo{year}{2013}): \emph{\bibinfo{title}{The two-handed tile assembly
  model is not intrinsically universal}}.
\newblock In: {\sl \bibinfo{booktitle}{ICALP: 40th International Colloquium on
  Automata, Languages and Programming. Proceedings, part 1}}, {\sl
  \bibinfo{series}{LNCS}} \bibinfo{volume}{7965}, \bibinfo{address}{Riga,
  Latvia}, pp. \bibinfo{pages}{400--412}, \doi{10.1007/978-3-642-39206-1\_34}.
\newblock \bibinfo{note}{Arxiv preprint
  \href{http://arxiv.org/abs/1306.6710}{\texttt{arXiv:1306.6710}} [cs.CG]}.

\bibitemdeclare{article}{dotysurvey}
\bibitem{dotysurvey}
\bibinfo{author}{David \surnamestart Doty\surnameend} (\bibinfo{year}{2012}):
  \emph{\bibinfo{title}{Theory of Algorithmic Self-Assembly}}.
\newblock {\sl \bibinfo{journal}{Communications of the ACM}}
  \bibinfo{volume}{55(12)}, pp. \bibinfo{pages}{78--88},
  \doi{10.1145/2380656.2380675}.

\bibitemdeclare{inproceedings}{IUSA}
\bibitem{IUSA}
\bibinfo{author}{David \surnamestart Doty\surnameend}, \bibinfo{author}{Jack~H.
  \surnamestart Lutz\surnameend}, \bibinfo{author}{Matthew~J. \surnamestart
  Patitz\surnameend}, \bibinfo{author}{Robert~T. \surnamestart
  Schweller\surnameend}, \bibinfo{author}{Scott~M. \surnamestart
  Summers\surnameend} \& \bibinfo{author}{Damien \surnamestart
  Woods\surnameend} (\bibinfo{year}{2012}): \emph{\bibinfo{title}{The tile
  assembly model is intrinsically universal}}.
\newblock In: {\sl \bibinfo{booktitle}{FOCS: Proceedings of the 53rd Annual
  IEEE Symposium on Foundations of Computer Science}}, pp.
  \bibinfo{pages}{439--446}, \doi{10.1109/FOCS.2012.76}.

\bibitemdeclare{inproceedings}{USA}
\bibitem{USA}
\bibinfo{author}{David \surnamestart Doty\surnameend}, \bibinfo{author}{Jack~H.
  \surnamestart Lutz\surnameend}, \bibinfo{author}{Matthew~J. \surnamestart
  Patitz\surnameend}, \bibinfo{author}{Scott~M. \surnamestart
  Summers\surnameend} \& \bibinfo{author}{Damien \surnamestart
  Woods\surnameend} (\bibinfo{year}{2009}): \emph{\bibinfo{title}{Intrinsic
  Universality in Self-Assembly}}.
\newblock In: {\sl \bibinfo{booktitle}{STACS: Proceedings of the 27th
  International Symposium on Theoretical Aspects of Computer Science}}, pp.
  \bibinfo{pages}{275--286}, \doi{10.4230/LIPIcs.STACS.2010.2461}.

\bibitemdeclare{techreport}{hendricks2013signal}
\bibitem{hendricks2013signal}
\bibinfo{author}{Jacob \surnamestart Hendricks\surnameend},
  \bibinfo{author}{Jennifer~E \surnamestart Padilla\surnameend},
  \bibinfo{author}{Matthew~J \surnamestart Patitz\surnameend} \&
  \bibinfo{author}{Trent~A \surnamestart Rogers\surnameend}
  (\bibinfo{year}{2013}): \emph{\bibinfo{title}{Signal Transmission Across Tile
  Assemblies: {3D} Static Tiles Simulate Active Self-Assembly by {2D}
  Signal-Passing Tiles}}.
\newblock \bibinfo{type}{Technical Report}.
\newblock \bibinfo{note}{Arxiv preprint
  \href{http://arxiv.org/abs/1306.5005}{\texttt{arXiv:1306.5005}} [cs.ET]}.

\bibitemdeclare{inproceedings}{HendricksPatitzTAMCA}
\bibitem{HendricksPatitzTAMCA}
\bibinfo{author}{Jacob \surnamestart Hendricks\surnameend} \&
  \bibinfo{author}{Matthew~J. \surnamestart Patitz\surnameend}:
  \emph{\bibinfo{title}{On the Equivalence of Cellular Automata and the Tile
  Assembly Model}}.
\newblock In: {\sl \bibinfo{booktitle}{Proceedings of Machines, Computations
  and Universality (MCU 2013)}}, \bibinfo{address}{University of Z\"{u}rich,
  Switzerland. September 9-12, 2013}.
\newblock \bibinfo{note}{To appear}.

\bibitemdeclare{techreport}{jonoska2012active}
\bibitem{jonoska2012active}
\bibinfo{author}{Natasha \surnamestart Jonoska\surnameend} \&
  \bibinfo{author}{Daria \surnamestart Karpenko\surnameend}
  (\bibinfo{year}{2012}): \emph{\bibinfo{title}{Active tile self-assembly,
  self-similar structures and recursion}}.
\newblock \bibinfo{type}{Technical Report}.
\newblock \bibinfo{note}{Arxiv preprint
  \href{http://arxiv.org/abs/1211.3085}{\texttt{arXiv:1211.3085}} [cs.ET]}.

\bibitemdeclare{incollection}{kari2012triangular}
\bibitem{kari2012triangular}
\bibinfo{author}{Lila \surnamestart Kari\surnameend},
  \bibinfo{author}{Shinnosuke \surnamestart Seki\surnameend} \&
  \bibinfo{author}{Zhi \surnamestart Xu\surnameend} (\bibinfo{year}{2012}):
  \emph{\bibinfo{title}{Triangular and hexagonal tile self-assembly systems}}.
\newblock In: {\sl \bibinfo{booktitle}{Computation, Physics and Beyond}},
  \bibinfo{publisher}{Springer}, pp. \bibinfo{pages}{357--375},
  \doi{10.1007/978-3-642-27654-5\_28}.

\bibitemdeclare{inproceedings}{LafitteW07}
\bibitem{LafitteW07}
\bibinfo{author}{Gr{\'e}gory \surnamestart Lafitte\surnameend} \&
  \bibinfo{author}{Michael \surnamestart Weiss\surnameend}
  (\bibinfo{year}{2007}): \emph{\bibinfo{title}{Universal Tilings}}.
\newblock In: {\sl \bibinfo{booktitle}{{STACS} 2007, 24th Annual Symposium on
  Theoretical Aspects of Computer Science, Aachen, Germany, February 22-24,
  2007, Proceedings}}, {\sl \bibinfo{series}{LNCS}} \bibinfo{volume}{4393},
  \bibinfo{publisher}{Springer}, pp. \bibinfo{pages}{367--380},
  \doi{10.1007/978-3-540-70918-3\_32}.

\bibitemdeclare{inproceedings}{LafitteW09}
\bibitem{LafitteW09}
\bibinfo{author}{Gr{\'e}gory \surnamestart Lafitte\surnameend} \&
  \bibinfo{author}{Michael \surnamestart Weiss\surnameend}
  (\bibinfo{year}{2009}): \emph{\bibinfo{title}{An Almost Totally Universal
  Tile Set}}.
\newblock In: {\sl \bibinfo{booktitle}{TAMC: Theory and Applications of Models
  of Computation, 6th Annual Conference, Changsha, China, May 18-22, 2009.
  Proceedings}}, {\sl \bibinfo{series}{LNCS}} \bibinfo{volume}{5532},
  \bibinfo{publisher}{Springer}, pp. \bibinfo{pages}{271--280},
  \doi{10.1007/978-3-642-02017-9\_30}.

\bibitemdeclare{techreport}{temp1notIU}
\bibitem{temp1notIU}
\bibinfo{author}{Pierre-\'{E}tienne \surnamestart Meunier\surnameend},
  \bibinfo{author}{Matthew~J. \surnamestart Patitz\surnameend},
  \bibinfo{author}{Scott~M. \surnamestart Summers\surnameend},
  \bibinfo{author}{Guillaume \surnamestart Theyssier\surnameend},
  \bibinfo{author}{Andrew \surnamestart Winslow\surnameend} \&
  \bibinfo{author}{Damien \surnamestart Woods\surnameend}
  (\bibinfo{year}{2013}): \emph{\bibinfo{title}{Intrinsic universality in tile
  self-assembly requires cooperation}}.
\newblock \bibinfo{type}{Technical Report}.
\newblock \bibinfo{note}{Arxiv preprint
  \href{http://arxiv.org/abs/1304.1679}{\texttt{arXiv:1304.1679}} [cs.CC]}.

\bibitemdeclare{inproceedings}{Ollinger08}
\bibitem{Ollinger08}
\bibinfo{author}{Nicolas \surnamestart Ollinger\surnameend}:
  \emph{\bibinfo{title}{Universalities in cellular automata a (short) survey}}.
\newblock In: {\sl \bibinfo{booktitle}{JAC: Symposium on Cellular Automata
  Journ\'ees Automates Cellulaires, 2008}}, pp. \bibinfo{pages}{102--118}.
\newblock
  \bibinfo{note}{\href{http://hal.archives-ouvertes.fr/hal-00271840}{\texttt{hal-00271840}}}.

\bibitemdeclare{article}{ollinger-fourstates}
\bibitem{ollinger-fourstates}
\bibinfo{author}{Nicolas \surnamestart Ollinger\surnameend} \&
  \bibinfo{author}{Gaétan \surnamestart Richard\surnameend}
  (\bibinfo{year}{2011}): \emph{\bibinfo{title}{Four states are enough!}}
\newblock {\sl \bibinfo{journal}{Theoretical Computer Science}}
  \bibinfo{volume}{412}(\bibinfo{number}{1-2}), pp. \bibinfo{pages}{22--32},
  \doi{10.1016/j.tcs.2010.08.018}.

\bibitemdeclare{article}{patitz2013introduction}
\bibitem{patitz2013introduction}
\bibinfo{author}{Matthew~J \surnamestart Patitz\surnameend}
  (\bibinfo{year}{2013}): \emph{\bibinfo{title}{An introduction to tile-based
  self-assembly and a survey of recent results}}.
\newblock {\sl \bibinfo{journal}{Natural Computing}}, pp.
  \bibinfo{pages}{1--30}, \doi{10.1007/s11047-013-9379-4}.

\bibitemdeclare{inproceedings}{RotWin00}
\bibitem{RotWin00}
\bibinfo{author}{Paul W.~K. \surnamestart Rothemund\surnameend} \&
  \bibinfo{author}{Erik \surnamestart Winfree\surnameend}
  (\bibinfo{year}{2000}): \emph{\bibinfo{title}{The Program-size Complexity of
  Self-Assembled Squares (extended abstract)}}.
\newblock In: {\sl \bibinfo{booktitle}{STOC '00: Proceedings of the
  thirty-second annual ACM Symposium on Theory of Computing}},
  \bibinfo{publisher}{ACM}, \bibinfo{address}{Portland, Oregon, United States},
  pp. \bibinfo{pages}{459--468}, \doi{10.1145/335305.335358}.

\bibitemdeclare{article}{SolWin07}
\bibitem{SolWin07}
\bibinfo{author}{David \surnamestart Soloveichik\surnameend} \&
  \bibinfo{author}{Erik \surnamestart Winfree\surnameend}
  (\bibinfo{year}{2007}): \emph{\bibinfo{title}{Complexity of Self-Assembled
  Shapes}}.
\newblock {\sl \bibinfo{journal}{SIAM Journal on Computing}}
  \bibinfo{volume}{36}(\bibinfo{number}{6}), pp. \bibinfo{pages}{1544--1569},
  \doi{10.1137/S0097539704446712}.

\bibitemdeclare{article}{Wang61}
\bibitem{Wang61}
\bibinfo{author}{Hao \surnamestart Wang\surnameend} (\bibinfo{year}{1961}):
  \emph{\bibinfo{title}{Proving Theorems by Pattern Recognition -- {II}}}.
\newblock {\sl \bibinfo{journal}{The Bell System Technical Journal}}
  \bibinfo{volume}{XL}(\bibinfo{number}{1}), pp. \bibinfo{pages}{1--41},
  \doi{10.1002/j.1538-7305.1961.tb03975.x}.

\bibitemdeclare{phdthesis}{Winf98}
\bibitem{Winf98}
\bibinfo{author}{Erik \surnamestart Winfree\surnameend} (\bibinfo{year}{1998}):
  \emph{\bibinfo{title}{Algorithmic Self-Assembly of {D}{N}{A}}}.
\newblock Ph.D. thesis, \bibinfo{school}{California Institute of Technology}.

\bibitemdeclare{inproceedings}{nubots}
\bibitem{nubots}
\bibinfo{author}{Damien \surnamestart Woods\surnameend},
  \bibinfo{author}{Ho-Lin \surnamestart Chen\surnameend},
  \bibinfo{author}{Scott \surnamestart Goodfriend\surnameend},
  \bibinfo{author}{Nadine \surnamestart Dabby\surnameend},
  \bibinfo{author}{Erik \surnamestart Winfree\surnameend} \&
  \bibinfo{author}{Peng \surnamestart Yin\surnameend} (\bibinfo{year}{2013}):
  \emph{\bibinfo{title}{Active self-assembly of algorithmic shapes and patterns
  in polylogarithmic time}}.
\newblock In: {\sl \bibinfo{booktitle}{ITCS: Proceedings of the 4th conference
  on Innovations in Theoretical Computer Science}},
  \bibinfo{organization}{ACM}, pp. \bibinfo{pages}{353--354},
  \doi{10.1145/2422436.2422476}.
\newblock \bibinfo{note}{Arxiv preprint
  \href{http://arxiv.org/abs/1301.2626}{\texttt{arXiv:1301.2626}} [cs.DS]}.

\end{thebibliography}
\end{document}